# A Machine Learning Framework for Biometric Authentication using Electrocardiogram

## Song-Kyoo Kim[1], Chan Yeob Yeun[1], Ernesto Damiani[1] and Nai-Wei Lo[2]

[1] Center for Cyber-Physical Systems, ECE Dept., Khalifa University, Abu Dhabi, United Arab Emirates
[2] Department of Information Management, National Taiwan University of Science and Technology, Taipei, Taiwan

Corresponding author: Chan Yeob Yeun (e-mail: chan.yeun@ku.ac.ae).

**ABSTRACT** This paper introduces a framework for how to appropriately adopt and adjust Machine Learning (ML) techniques used to construct Electrocardiogram (ECG) based biometric authentication schemes. The proposed framework can help investigators and developers on ECG based biometric authentication mechanisms define the boundaries of required datasets and get training data with good quality. To determine the boundaries of datasets, use case analysis is adopted. Based on various application scenarios on ECG based authentication, three distinct use cases (or authentication categories) are developed. With more qualified training data given to corresponding machine learning schemes, the precision on ML-based ECG biometric authentication mechanisms is increased in consequence. ECG time slicing technique with the R-peak anchoring is utilized in this framework to acquire ML training data with good quality. In the proposed framework four new measure metrics are introduced to evaluate the quality of ML training and testing data. In addition, a Matlab toolbox, containing all proposed mechanisms, metrics and sample data with demonstrations using various ML techniques, is developed and made publicly available for further investigation. For developing ML-based ECG biometric authentication, the proposed framework can guide researchers to prepare the proper ML setups and the ML training datasets along with three identified user case scenarios. For researchers adopting ML techniques to design new schemes in other research domains, the proposed framework is still useful for generating ML-based training and testing datasets with good quality and utilizing new measure metrics.

## I. INTRODUCTION

Because most application systems support Internet access for general users, identifying persons with their own body has become the trend for users to access application systems. In consequence, biometric authentication has become a hot research topic in recent years. Among various biometric authentication schemes such as fingerprint scanning and facial recognition, electrocardiogram authentication has the advantage of adopting live user body signals during authentication. In general, machine learning techniques are adopted to construct a verification model for user identification by getting user's live ECG data. Recently there are a number of state-of-art literatures on ECG based biometrics [1-4]. However, several ECG biometrics challenges still require further investigation such as authentication categorization, pre-processing for data quality enhancement, data acquisitions, selection on Deep Learning (DL) and other Machine Learning classification approaches [5].

This article introduces a ML framework for ECG based biometric authentication in order to mitigate identified challenges on ECG authentication. To better understand potential application environments for ECG authentication, it is necessary to identify basic application scenarios through use cases. In the proposed framework application scenarios using ECG authentication are categorized into three general use cases: Hospital (HOS), Security Check (SCK) and Wearable Devices (WD). Furthermore, new data pre-processing techniques including the baseline adjustment of frequency artifacts in the ECG, the ECG data noise removal technique for Power Line Interference (PLI), and flipping mechanism for ECG signal due to the wrong placement of electrodes, are proposed. In addition, time slicing technique are introduced in the framework to prepare ML-based training datasets along with new measure metrics developed for authentication precision evaluation. Four new measure metrics for data quality are introduced in the proposed framework. They are Mean Absolute Error Rate (MAER), Upper/Lower Range Control Limits (UCL/LCL), Accuracy Percentage within Ranges (APR), and Accuracy per UCL (APU).

Figure 1 illustrates an overview of the new framework model for ML-based ECG biometric authentication. Within the core process portion, several ML techniques are adopted: Decision Tree (DT) and Support Vector Machine (SVM) for regression approach, and Artificial Neural Network (ANN) and Convolutional Neural Network (CNN) for classification approach. In addition, time slicing technique for ECG data is developed and associated with the core process.

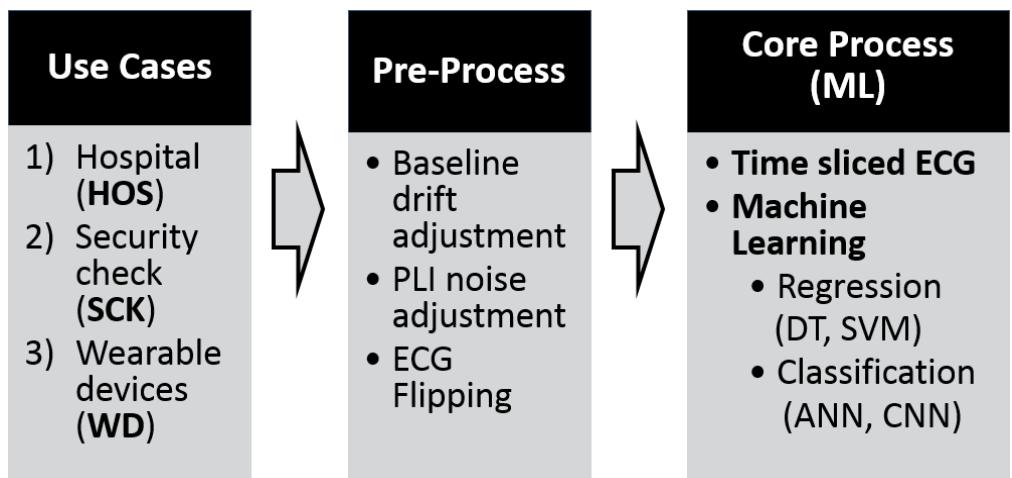

FIGURE 1. Overview of the New Framework Model for ECG based Biometric Authentication

Figure 2 demonstrates the framework processing flow for ECG biometric authentication using Machine Learning techniques. The whole proposed framework starts with selecting a proper user category for the target system environment and then moves onto the training phase to acquire corresponding ECG data from target users as the training dataset. Once the training data are received, the pre-process techniques are used on top of these data to get filtered ones with higher quality (less noise). The filtered data is used as the input of one of selected core process mechanisms shown in Figure 1 to generate an authentication evaluation model. The proposed core process currently supports both ECG reference database and trained Neural Network (NN) reference engine as the evaluation model for any ML-based ECG authentication mechanism to use them. The reference database (or the NN Engine) is generated when the training phase is completed. In the testing phase, an ECG based user authentication request associated with newly received ECG data is generated and the ECG data need to adopt data pre-process techniques to obtain filtered data with higher quality (less noise) first. Then the filtered data are sent to validation process as the input to check with the reference database (or the NN Engine) for the final decision on this user authentication request.

In addition, the new Matlab toolbox with this paper contains all proposed mechanisms, metrics and sample data with demonstrations using various ML techniques. This toolbox has been developed and publicly available for further investigation.

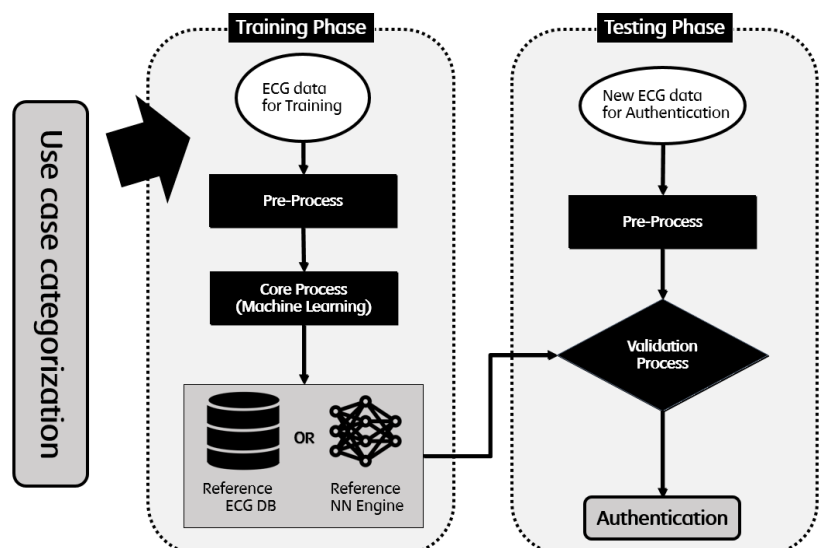

**FIGURE 2.** The Framework Processing Flow for EGC Biometric Authentication using Machine Learning Technologies

This article consists of seven sections. Section II addresses three new authentication categories based on use cases. Section III depicts three new pre-processing technologies for enhancing collected ECG data quality. In Section IV, we introduce the time slicing technique for ECG data to improve the training results of applied ML schemes and two approaches for ML data training. Section V presents four new data quality metrics. In Section VI, we describe the functions of our Matlab toolbox, which is used to evaluate the applicability of our proposed framework. Finally, a summary of the proposed framework and our contributions are given in Section VII.

## II. AUTHENTICATION CATEGORIZATION BASED ON USE CASES

Biometric authentication using ECG data has been widely studied [6-11]. Although there are a lot of literatures for ECG based authentication, all of them apply ECG based authentication schemes under different user environments and various ECG detection devices. Researchers with medical engineering expertise usually set up an electrocardiograph for collecting ECG data [6-8]. In contrast, researchers with electrical engineering expertise usually setup simple ECG sensors (usually embedded in wearable devices) for receiving ECG data [9-11]. Therefore, it is important to consider and understand user environment and what type of ECG detection device may be in use before developing an ECG based biometric authentication scheme.

A use case is a written description of how users will interact with the target system. Use case analysis is able to identify system requirements during the design stage and clarify critical information for system processes [12]. Possible use cases could be categorized through use case analysis. By adopting use case analysis technique on possible application scenarios for ECG based user authentication, three authentication categories are identified: patients in a hospital, identity check for people at a building entrance, and continuous authentication for personal usage (a.k.a. HOS, SCK and WD as shown in Figure 1). The corresponding user environment and assumptions for each category are addressed as follows. Notice that system performance requirement in terms of authentication speed and accuracy rate for each category is different and dependent on target application systems.

The description and the conditions of each category are explained in detail in the following section. All authentication use cases are similar but the performance measure schemes shall be different.

### A. AUTHENTICATION OF PATIENTS IN A HOSPITAL (HOS)
Traditionally, a patient will take an ECG test to diagnose whether a heart disease or a heart stress is occurred. The equipment for gathering ECG signals from a patient is usually elaborated and complicated for gathering the high quality for medical diagnostics. Therefore, the sampling time for getting ECG data is relatively long (from couple of minutes to hours dependent on the type of ECG test) and multiple leads are used during an ECG test. A new use case for ECG test is to identify patients in a hospital (Category 1; HOS use case). The assumption is that those patients have to register their identities (i.e., their names or legal identity numbers) along with their historical ECG data in advance. In addition, it is assumed that the measured ECG signals from the same patient are stable enough (i.e., the measured ECG signal values within a normal range) for both registration (training) and verification (testing) phases of an ECG based authentication scheme. Then the hospital can identify those patients with ECG based biometric authentication scheme next time the patients enter the hospital. Notice that a well-trained ECG user authentication model (or scheme) can identify a patient by evaluating live ECG signals in a very short period of time (less than a couple of seconds) [13] in comparison with questioning for the patient's name and his/her legal identity number by a nurse (it may take a couple of minutes). For patients losing consciousness in an emergency room, ECG based user (or patient) authentication can easily identify those patients. In general, a patient authentication in hospitals may be one of the major application environments for ECG based authentication schemes. This HOS use case is the most widely applied research environment in healthcare and medical industry [14]. There are a lot of public available databases containing historical ECG data such as the PhysioBank database [15].

Because of complexity for data collecting operations, the received ECG data could be flipped or contain certain noises caused by PLI or wrong position placement of electrodes. Therefore, it is necessary to perform data pre-processing mechanisms onto these ECG data to further improve data quality before using them to train evaluation model for user (or patient) authentication purpose.

### B. AUTHENTICATION OF PEOPLE AT A BUILDING ENTRANCE (SCK)

The second user authentication use case based on user ECG data is applied to security check (Category 2; SCK use case) for building entrance and room entrance if necessary. Most of companies have security check points to identify employees and visitors. With the availability of portable ECG detection devices or ECG detection sensors, ECG based biometric authentication system will become one of the user authentication choices for security check points among fingerprint scanning, facial recognition, voice identification, iris recognition, and retina scan. In this SCK use case, an ECG based authentication system can be used to identify registered regular employees and unknown persons (including visitors). The assumption is that legal employees have registered their identities (i.e., their names or legal identity numbers) along with their historical ECG data to this ECG authentication system in advance. In addition, it is assumed that the measured ECG signals from the same employee are stable enough for both registration and verification phases.

### C. CONTINUOUS AUTHENTICATION FOR PERSONAL USAGE (WD)

The third ECG based user authentication use case is for personal wearable devices (e.g., a smart watch) embedded with ECG sensors to continuously monitor whether the person currently wearing the wearable device is the genuine owner (Category 3; WD use case). In general, a wearable device only needs to authenticate that its user is its owner continuously. Since the heart beat period (R-R peak period) and the amplitude of ECG signals of a person may change dramatically when the person is under different body status such as walking, running, and sleeping, a new framing technique may be required to normalize the received ECG signals by filtering out potential signal noise caused by user body status. With an embedded ECG sensor and an ECG authentication module, a wearable device can also support the second factor authentication functionality for its user. By adopting two factor authentication scheme (e.g. authentication with both user password and user ECG data), a WD user can have more security control on his/her wearable device.

The summary of the use case categories is shown in Table I. Each use case category requires different measure metrics, different boundaries of person identification and different characteristics of referenced ECG data. As we can see in Table I, the HOS case does not consider the situation of patients with unknown identifications, whereas the authentication system in the SCK case has to define a process to handle the situation of unknown identifications occurred. The WD case must consider how to neutralize the ECG data noise because of different body status even from the same person.

TABLE I
DATASET BOUNDARIES CATEGORIZED BY AUTHENTICATION USE CASES

| Cat No. | Cat. Name | Known ID classification | Unknown ID | Personal Status |
|---|---|---|---|---|
| 1 | Hospital (HOS) | O | X | X |
| 2 | Security Check (SCK) | O | O | X |
| 3 | Wearable Devices (WD) | X* | O | O |

\* There is only one known ID in the WD case

Researchers may consider additional factors based on the use case categories. For instance the ECG sampling time for security check point should be shorter than the ECG sampling time for patients in a hospital. In the WD case, the sampling frequency for ECG signals should be much lower than the ECG sampling frequency generated by traditional medical measurement equipment on ECG signals because the objectives for utilizing collected ECG data are different. In addition, human operation errors such as misplacement of leads shall not be considered in WD category since different types of ECG devices are utilized between the HOS case and WD case.

## III. PRE-PROCESS FOR ECG DATA QUALITY ENHANCEMENT

The pre-process is about adjusting data before starting the core process (i.e., machine learning process). The signal processing techniques have been widely applied into adjusting ECG data since ECG data could be considered as signals. Many applications in the signal processing including the filter designs [16-18] and the Fourier Transforms [19-20] are applied for enhancing the ECG recognition. Although many pre-processes are adopted for enhancing signals, three process are recommended for enhancing ECG data before starting the machine learning process.

### A. BASELINE ADJUSTMENT

The baseline drift (also called the baseline wander) is a low frequency artifact in the ECG that arises from breathing, electrically charged electrodes [21]. Baseline adjustment process is removing the baseline wander. Typically, a complete baseline wander removal requires that the cut-off frequency of the high-pass filter be set higher than the lowest frequency in the signal. The majority of baseline wander removal techniques have in common that they cancel the low frequency components of the signal. The frequency of the baseline wander high-pass filter is usually set slightly below 0.5 Hz [22]. Although, these techniques are well studied and widely used [23], the proper frequency for the baseline drift removal should be determined in advance. During ECG data gathering, the baseline drift could be also occurred by certain movement of an applicant besides of the low frequency noise. Therefore, these filtering techniques may not useful if we do not know proper indicator of the cut-off frequency or expecting certain movement of an applicant (HOS case).

Alternatively, a curve fitting technique could be used of the baseline adjustment to avoid these problems. Curve fitting is the process of constructing a curve which finds the best fit to a series of data points [24] and polynomial curve fitting finds an exact fit to the data more smoothly [25]. The ECG baseline could be adjusted by subtracting the fitted curve data from the original ECG data in Figure 3.

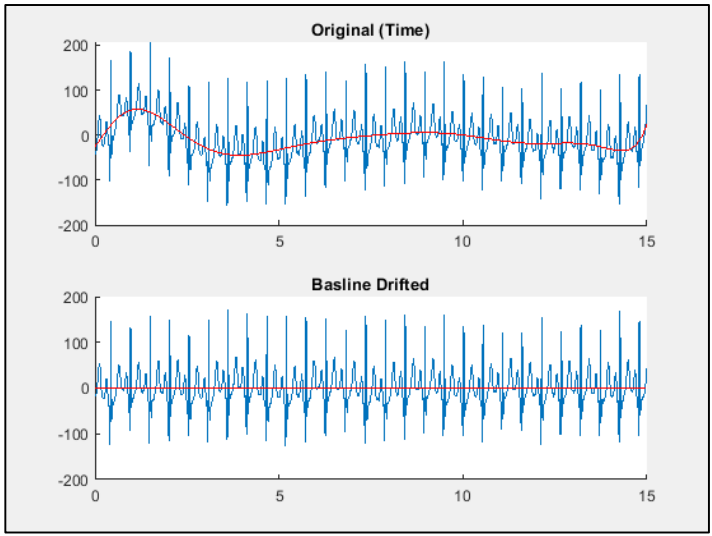

**FIGURE 3.** Baseline adjustment by using the polynomial curve fitting [15]

Typically, this baseline adjustment technique with the polynomial curve fitting is useful if the noise by low frequencies is not only problems of the baseline wander. In addition, it is automatically adjusting the baseline to zero (0) which avoids any ECG amplitude shifting. Atypical cause of the baseline wander except for a low frequency noise is certain movements of an applicant during ECG data gathering.

### B. POWER LINE INTERFERENCE NOISE REMOVAL
There are several types of noise signals including the baseline wander and the noise from the PLI which is coupled to signal carrying cables is particularly troublesome in medical equipment is also commonly happened in a hospital (HOS case) when cables carrying signals from the examination room to the monitoring equipment are prone to electromagnetic interference (EMI) of frequency by seamless supply lines [26]. Electromagnetic fields caused by a power line represent a common noise source in the ECG that is characterized by 50 or 60 Hz sinusoidal interference, possibly accompanied by a number of harmonics. Such narrow band noise cause problems interpreting low amplitude waves because it introduces unreliable and spurious waveforms [27]. The Infinite Impulse Response (IIR) notch filter is widely applied to remove PLI noise [28]. A notch filter rejects or attenuates signals in a specific frequency band called the stop band frequency range and pass the signals above and below this band [29]. These types of filters could be applied to remove the baseline wander but a proper target frequency should be determined before applying the filters. In the other hand, removing abnormal peak points in the frequency domain provides similar effects by using a notch filter without determining the target frequency in advance. Since PLI has the high peaks in the frequency domain, this technique could applied to remove PLI noise in Figure 4.

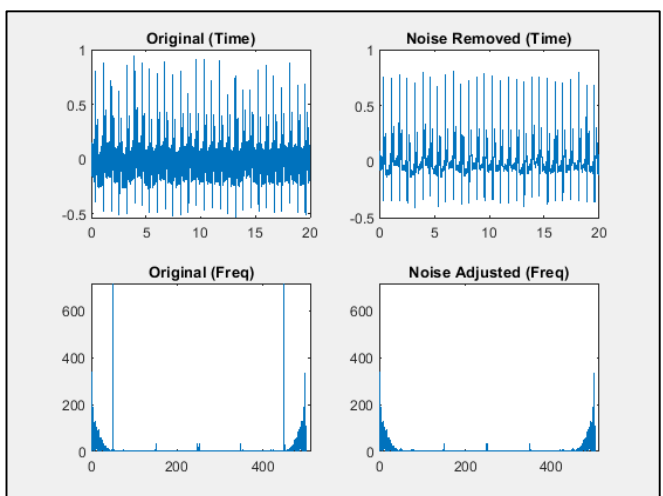

**FIGURE 4.** Noise adjustment for removing PLI frequency [30]

The Fourier Transform has been used to find abnormal peak points in the frequency domain. The abnormal peak points (typically 50-150 times higher than an average magnitude) in the frequency domain including PLI frequencies could be considered as a noise and these peak points are subject to remove for enhancing the original ECG data. Alternatively, the specific frequencies which occur noises could be removed. As it mentioned above, the 50 Hz frequency occurs the PLI and the 1 Hz frequency occurs the baseline wander. The ECG data could be improved by removing these frequencies by using the Fourier Transform even without using filters.

It is noted that the PLI typically happens in the HOS case but it would not happen in the WD case. Therefore, the PLI noise removal process may not be applied in the wearable device authentication because the PLI is happened only if they use a medical ECG equipment even for the SCK case. Basically, the pre-process for PLI noise removal shall not be applicable for an ECG receiver which uses a simple sensor type device.

### C. FLIPPING SIGNAL

The medical machines including an electrocardiograph typically requires professionals to setup and measure ECG signals. But even professionals make mistakes including wrong placement of electrodes. Flipping ECG signal is atypical situation by mistakes of operators in hospitals. Therefore, it is better to check whether a target ECG data (either training or testing phases) is flipped or not and Fig 5 shows an example of ECG flipping. This pre-process adjusts the flipped ECG data if an original data is flipped (if it is not, an adjusted data is same as original data). The simplified way of flipping ECG data is determining the flipping status by checking the mean of small portions of the data instead of checking a whole data.

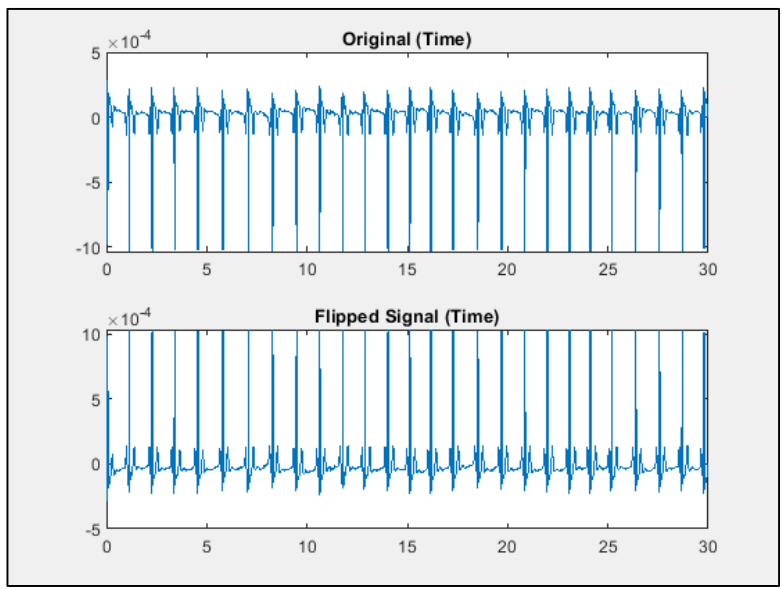

**FIGURE 5.** Flipping ECG data example [31]

It is not mandatory to apply all three pre-processes at once rather, it depends on the authentication use cases. For instance, flipping signal shall not be applicable for wearable devices (the WD case). On the other hand, other additional processes such as changing sampling frequencies for registered ECG data (a reference database) and changing the unit of ECG amplitude are also considered as a pre-process (specially, system upgrade for the company entry system). Once required pre-processes are completed, the core process which is mainly the ECG time slicing and the machine learning training shall be started and the following section provides the details.

### IV. TIME SLICING AND MACHINE LEARNING

The time slicing technique is considered for building up the dataset of the machine learning training. This approach is especially applicable for building up the machine learning training data. The ECG data are sliced based on a slice (window) time (typically known as a sliding window) with the R-peak anchoring. This method could generate enough data samples and each sliced data becomes a sample input for the machine learning training. The time sliced ECG dataset is very flexible not only to mix with other training inputs but also to apply various ML training methods which is explained in the subsection B.

## A. TIME SLICED ECG DATA FOR MACHINE LEARNING

The QRS complex is the combination of three of the graphical deflections seen on a typical ECG which is usually the central and most visually obvious part of the tracing [32]. An R-peak which is the maximum of a QRS complex. It indicates one heartbeat and the moment of the R-peak is commonly used for the anchor of the QRS complex including the R-peak detection and optimizing the sliding window time [33]. Time slicing, which is basically slicing ECG data in the time domain, is targeted for chopping the ECG signal from an R-peak moment to the sliding window period and layering these pieces based on the R-peak moment (i.e., R-peak anchoring). Each slicing peace based on the R-peak anchoring in Figure 6 becomes sample inputs for the machine learning training. In this paper, the average minimum of a heartbeat interval from atypical heart rate [34] is chosen as a slicing time (i.e., sliding window) which is 0.6 seconds which is equivalent with 100 bps heartbeat rate. The optimized sliced window time depends on the purpose of ECG based projects and some performance measures for the machine learning are somewhat depending on the ECG slicing time. This optimization problem is not considered at this moment but optimizing window time for biometrics purposes are another interesting ECG based security research topics.

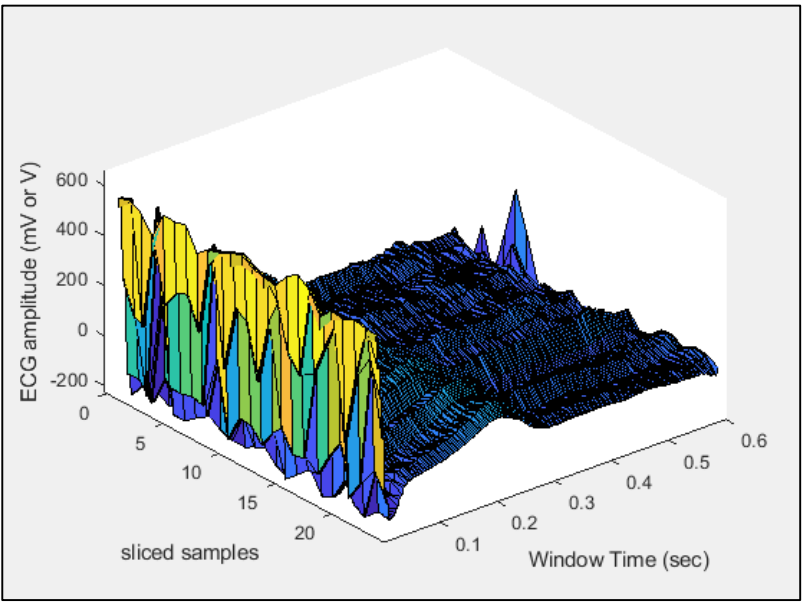

**FIGURE 6.** ECG Time Slicing with the R-peak anchoring [15]

## B. MACHINE LEARNING PROCESS FOR TRAINING DATA

Machine learning is a subset of artificial intelligence (AI) which computer systems use to effectively perform a specific task without using explicit instructions. Machine learning algorithms build a mathematical model of training data to make predictions (regressions) or decisions (classification/ pattern recognition) without being explicitly programmed [35]. It has been widely studied for analyzing ECG data by using AI techniques [5, 6, 8, 36-39].

This research focuses to train the time sliced ECG data by using ML techniques. Various techniques are provided to demonstrate how a machine learning model is applied into the ECG based biometric authentication.

### 1) MACHINE LEARNING FOR REGRESSION APPROACH

A multi-variable regression is a data mining task of predicting a value of the target by building a model based on multiple variables. The Support Vector Machine (SVM) is a discriminative classifier formally defined by a separating pattern [40]. This technique has been applied for the regression since 1990s [41]. The Decision Tree (DT) could also build regression model in the forms of a tree structure. It breaks down a dataset into smaller subsets while at the same time an associated decision tree is incremented [42-43]. It is shown that the DT technique delivers the better performance compare to the SVM technique for building the regression model in Table II based on the dataset from the PhysioBank database [15].

TABLE II
COMPARISON TABLE OF DT AND SVM

| DT | | SVM | |
|---|---|---|---|
| Results | | Results | |
| RMSE | 33.761 | RMSE | 36.168 |
| R-Squared | 0.59 | R-Squared | 0.54 |
| MSE | 1139.8 | MSE | 1308.1 |
| MAE | 19.164 | MAE | 19.01 |
| Prediction speed | ~740000 obs/sec | Prediction speed | ~600 obs/sec |
| Training time | 5.2194 sec | Training time | 2008.2 sec |

Besides of these two ML techniques, other several machine learning models could be applied for developing a regression model. In addition, choosing a DT model might be the best option only dedicated for a particular setup of computer systems and particular (time sliced) ECG datasets. The optimal selection of a proper ML model for a regression approach might vary by a different dataset or by a different computing system.

### 2) MACHINE LEARNING FOR CLASSIFICATION APPROACH

Classification is target to identify the category set based on the basis of a training set of data containing observations [44]. The set of the identification could be considered as a set of categories and each identification becomes each category set. Neural network (NN) models (either Artificial Neural Networks [45] or Convolutional Neural Networks [46]) could be designed by using the time sliced ECG dataset as the inputs (see Figure 7). The performance of network models shall not be the same and designing a proper NN model is yet another research topics [47-50]. Regardless which NN models are applied, both neural network models are compatible to use the same input nodes and the same output nodes. Therefore, the time sliced ECG dataset based on the R-peak anchoring could be the input samples for both neural network models.

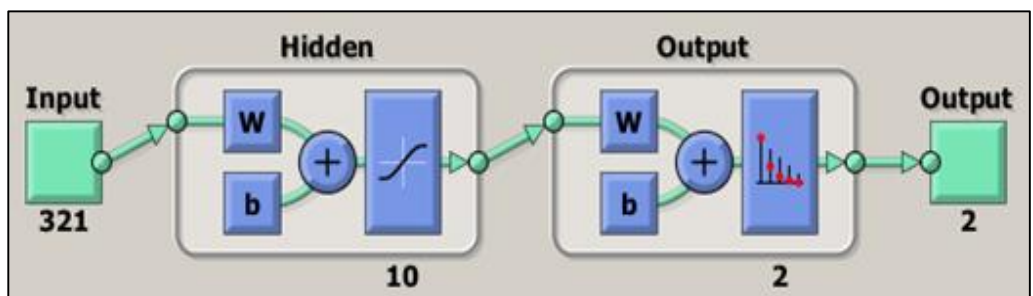

**FIGURE 7.** Design of 10 hidden layer Convolution Neural Network (WD)

It is noted that this paper does not deal with any particular models and the paper is only targeted to provide the guidelines to design the better NN for authentication projects.

The following section explains the data quality measures to check the quality of both a machine learning models and training dataset.

## V. DATA QUALITY MEASURES

Evaluating machine learning algorithms is an essential part of any ML projects [51] and delivering the good quality samples is vital for the ML algorithm evaluations. Some performance measures (evaluation metrics) are the values of evaluation results of sample data qualities and regression approaches based authentication systems [52]. Typical data quality measures and performance measures for a regression approach are as follows:

- Sum of Squares Error
- Sum of Squares Total
- Mean Squared Error
- Mean Absolute Error

Besides of typical quality measures, additional new measurement factors are introduced in the paper. These quality measures are not only applied for evaluating regression approached machine learning systems but also applied for the criteria of validating samples. The new data quality measures are as follows:

- Mean Absolute Error Rate
- Upper and Lower Range Control Limits
- Accuracy Percentage within Ranges
- Accuracy per Upper Control Limit

The details of above new data quality measures are explained on the following subsections.

### A. MEAN ABSOLUTE ERROR RATE (MAER)

The Mean Square Error (MSE) and the root MSE (RMSE) are useful for measuring errors for evaluating a machine learning model. The MSE is an average squared difference between the estimated two datasets [53]. Similarly, the RMSE is the square root of the MSE and both measures are widely used for machine learning evaluation metrics. Although both (MSE, RMSE) measures are commonly used for most of regression based machine learning model evaluations, both measures are hard to realize whether the model is good or bad by looking themselves. A new measure to evaluate either machine learning engines or training dataset is introduced in this paper. The Mean Absolute Error Rate (MAER) is defined as follows:

$$MAER = \frac{1}{N}\sum_{n=1}^{N} \frac{|Y_n - \mu_n|}{\mu_n + \epsilon}, \epsilon \sim 0 \tag{1}$$

where $\mu_n$ is the reference (typically the mean) value of the data $Y_n$. It is noted that the formula (1) has been designed to avoid for dividing zero. The reference values ($\mu_n, n = 1, \ldots, N$) in the MAER are the mean values of the prediction after the machine learning training. Unlike MAD (Mean Absolute Difference) and MAE (Mean Absolute Error), the MAER is calculated based on the moving mean (i.e., the reference values) instead of using a fixed mean or arbitrary differences. It provides the portion of the range based on the reference values and it is a useful metric for measuring the regression performance. The value of the MAER is always going to be between 0 (zero) and ∞ and a smaller value indicates a better performance.

The MAER could be used not only for evaluating the quality of the ML training samples but also for evaluating a regression approached machine learning model. The range values of the data which are greater than the MAER threshold is considered as outliers and these particular data could be rejected for improving the data quality. Finding the proper MAER value as a threshold for data quality is not a simple task but a quality engineering technique (i.e., Statistical Process Control) could help to solving this problem in the subsection B)

### B. UPPER/LOWER RANGE CONTROL LIMITS (UCL, LCL)

Statistical process control (SPC) is defined as the use of statistical techniques to control a process production method in the quality engineering and SPC is often used interchangeably with statistical quality control [54]. Control charts and acceptance sampling to quality control are main applications of SPC [55]. The control charts has two types: one is charts about the mean of the data (X-charts) and the other is charts about the range (a variance or a standard deviation) of the data (R-charts). Control limits (*Upper Control Limit* and *Lower Control Limit*) are horizontal lines drawn on a control chart, usually at a distance of ±3 standard deviations (or ±6 standard deviations) of the plotted statistic from the mean of data. The R-charts could be built not only with the reference values but also without the reference values. Both charts shows the quality of the data and the control values could be used for the criteria of rejecting outliers before training the data. The control limits (UCL and LCL) of the range values R could be found as follows:

$$UCL = \bar{R}_Y + \sigma(b)\bar{R}_Y, \tag{2}$$

$$LCL = Max(0, \bar{R}_Y - \sigma(b)\bar{R}_Y), \tag{3}$$

where $\sigma(b)$ is the portion of the sigma level in the normal distribution for accepting the data based on the ranges which is typically 3 sigma (i.e., $b = 3$; $\sigma(b) = \Phi(b) - \Phi(0)$; $\Phi(z)$ is the normal distribution function). The *R bar* is the mean of the range values. Both square errors (MSE) and absolute errors (MAER, MAE) could be used as the set of range values for determining the control limits. The smaller UCL indicates a better performance and the UCL becomes the threshold range value for evaluating the datasets.

Basically, the R-chart from the SPC provides the threshold for organizing proper datasets from the raw data in both training and testing phases.

### C. ACCURACY PERCENTAGE WITHIN RANGES (APR)

The *Accuracy Percentage within Ranges* (*APR*) is the portion of the ECG data within the ranges between 0 and UCL. It is the counting numbers within the ranges out of the total number of the sliced ECG data samples.

$$APR = \frac{n(\{R \leq UCL\})}{n(\Omega_R)} \quad (4)$$

The *APR* indicates the quality of the time sliced ECG data even before validating data and the larger *APR* indicate the better performance. This value also becomes the threshold for rejecting the testing ECG data before the comparison with the reference data. The bigger *APR* values indicate the better quality of both datasets and ML systems.

### D. ACCURACY PER UCL (APU)

The *Accuracy percentage within ranges per the upper control limit* (*APU*) is the ratio between the *APR* and the UCL which shows how much the training has been improved. The *APR* could be small even the accuracy is high when the UCL is getting smaller at the same time. The *APU* is calculated as follows:

$$APU = \frac{APR}{UCL} \quad (5)$$

The *APU* is another performance indicator to understand the quality of data and a machine learning system.

Existing quality measures for ML datasets are still applicable and these measures could be used selectively. The new data quality measures in the paper could be alternative evaluation metrics for regression based Machine Learning systems for biometric authentication in general. The selection of proper quality measures will be the matter of choice which depends on the use case category and the purpose of a project. The proposed new measures are adopting the techniques from a different engineering fields and these measures are efficient for measuring the quality of both training and testing datasets.

## VI. MATLAB TOOLBOX

The previous sections from I-IV provide the core process with the fundamental components for adopting ML techniques in the ECG based biometric authentication systems. The proposed Toolbox for demonstrating the proposed processes and techniques in each section is actually implemented as the functions of the Matlab. The *amgecg Toolbox* (Amang ECG Toolbox) is the set of the Matlab functions and researchers could use their own ECG authentication projects and this section introduces some of functions in this Toolbox.

### A. TOOLS FOR PRE-PROCESS AND TIME SLICING

Three processes which include the baseline drift adjustment in Figure 3, the noise adjustments in Figure 4 and the flipping ECG data in Figure 5 have been newly introduced as the pre-process of the Machine Learning adaptations and the Matlab functions in the new Toolbox are as follows:

- *baselinedrift*
- *enhancednoiseadjust*
- *pseudoflip*

The way of using each function could be found by using "help" function of the Matlab. Users can earn the detailed information for how to use each function via the Matlab command window as shown in Figure 8.

```
Command Window
>> help baselinedrift
  ECG baseline drift correction

  Usage:
    Adjusted_Dat = baselinedrift(sfq, ECG_data, DispOpn)
  Output:
    Ajusted_ECG : Adjusted ECG signals
  Input:
       sfq     : Given sampling frequency [Hz]
    ECG_data   : Original ECG data to be adjusted [mV]
    DispOpn    : Display Option [Off (0) or On (1)]
  Note:
    - Required Matlab file(s): polyfit
    - Baseline adjustment by using Polynomial Curve Fitting
    - Ploting the comparison between Orignal and Adjusted data
    - Default order of coefficient (cof) = 10

Made by Amang Kim [v0.25 || 3/1/2019]
Package of amgecg (Amang ECG) Toolbox [Rel. Ver. 0.6 || 3/5/2019]

>>
```

**FIGURE 8.** Using 'help' function in the Matlab command window

The function of the time slicing of the ECG data also provided in the *amgecg Toolbox* as shown in Figure 6 and this *timeslicedecg* function is one of core for adopting ML techniques into ECG authentication projects and it gives flexibilities to adopt various ML techniques for ECG data. The ML system could be designed by using time sliced ECG data and each sliced data is considered as the input of any ML systems. It is noted that actual Machine Learning implementations such as the DT regression and CNN classification (Figure 7) are not included in the *amgecg Toolbox* but users can implement ML systems by using this Toolbox.

### B. TOOLS FOR DATA QUALITY
The *amgecg Toolbox* also contains the functions for analyzing the data quality. The indicators of the data quality that have been mentioned in Section V are provided in the toolbox and the related functions are as follows:

- *rangecontrol*
- *maer*
- *mseamg*
- *maerdataqualityengine*
- *msedataqualityengine*

The functions for data quality are combination of some basic and integrated functions. Users can find the details of each function on the Matlab "help" function. In addition, the demo file is also available with the Toolbox as shown in Figure 9.

VOLUME XX, 2017

```
Command Window
Loading the record of C:\AMG_Lounge\Mathwork_AMG\AMG_Pub\amgecgdb\205.mat between 0 [sec] and 20 [sec]....
Finding the frequency is 400 [Hz] ......
Starting time is revised from 0 [sec] to 20 [sec]..........
Generating the adjust signal by using Polynomial curve fitting with 10 orders.....
The ECG signal is NOT flipped ..................
The ECG signal might be flipped and Try to do the baseline adjustment before flipping..........
Finding total 30 R-peaks on the signal........

DQ3 =

  struct with fields:

                DBPath: 'C:\AMG_Lounge\Mathwork_AMG\AMG_Pub\amgecgdb\'
                  Freq: 400
            SampleTime: [0 20]
             SliceTime: [0 1]
            NumofSlice: 29
           NumofRecord: 10
            FileRecord: [101 102 103 104 105 201 202 203 204 205]
            SigmaLevel: 3
                  MAER: [1×10 double]
    min_mean_Max_MAER_APR: [0.7917 0.8821 1]
    min_mean_Max_MAER_APU: [4.6800e-06 3.2677e-04 9.0906e-04]
    min_mean_Max_MAER_UCL: [1.1000e+03 2.2191e+04 1.8454e+05]
            FullReport: [10×4 double]

Accepted =

   101   102   103   104   201   202   203   204   205

Rejected =

   105

>>
```

**FIGURE 9.** Demonstration of MAER based Data Quality Analysis

It is noted that the Matlab source codes (i.e., *amgecg Toolbox*) are available on the GitHub[1] and readers can freely use them. In addition, the demos of using the functions of the *amgecg Toolbox* could be found on the YouTube[2].

## VII. CONCLUSION

As new ECG detection devices become portable, lightweight, embeddable with smartphones and wearable devices, and connectable with remote servers through wireless technologies in the near future, ECG based biometric authentication will be deployed on massive application systems all over the world. To get high accuracy on user authentication, ML techniques are generally adopted to build a more robust evaluation model for ECG based biometric authentication. In this paper a generalized machine learning framework for ECG based biometric authentication is introduced. The proposed framework describes the general data processing flow of a ML-based ECG authentication mechanism along with various function features to help researchers easily design and evaluate a ML-based ECG user authentication scheme. Those functions include three general authentication categories for ECG user authentication, three new data pre-processing techniques, a time slicing technique to generate high quality ECG datasets, four new data quality metrics, and a publicly available Matlab Toolbox (i.e., *amgecg Toolbox*). For people using ML technologies to investigate other topics instead of ECG based biometric authentication, several data pre-processing techniques and newly defined measure metrics offered by the proposed framework are still useful and can help researchers accelerate the development of their ML-based schemes.

## ACKNOWLEDGMENT
The authors would like to thank A. Khandoker and H. Alsafar for providing some ECG datasets that have been used in the article. Special thanks go to Jiankun Hu for his invaluable comments of this new ML adaptation process for ECG biometrics. The authors gratefully acknowledge supports from the Center for Cyber-Physical Systems in Khalifa University, under the Grant Number 8474000137-RC1-C2PS-T3 and also from the Taiwan Information Security Center (TWISC) and Ministry of Science and Technology, Taiwan, under the Grant Number MOST 107-2218-E-011-002.

---

[1] https://github.com/amangkim/amgecg_toolbox
[2] http://youtu.be/texyM7Gzz3c